# Architecting Safety Supervisors for High Levels of Automated Driving

Martin Törngren*, Xinhai Zhang*, Naveen Mohan*, Matthias Becker*†, Lars Svensson*, Xin Tao*, De-Jiu Chen*, Jonas Westman*

*Department of Machine Design
†Department of Electronics and Embedded Systems
Kungliga Tekniska Högskolan (KTH), Stockholm, Sweden
{martint, xinhai, naveenm, mabecker, larsvens, taoxin, chendj, jowestm}@kth.se

*Abstract*— The complexity of automated driving poses challenges for providing safety assurance. Focusing on the architecting of an Autonomous Driving Intelligence (ADI), i.e. the computational intelligence, sensors and communication needed for high levels of automated driving, we investigate so called safety supervisors that complement the nominal functionality. We present a problem formulation and a functional architecture of a fault-tolerant ADI that encompasses a nominal and a safety supervisor channel. We then discuss the sources of hazardous events, the division of responsibilities among the channels, and when the supervisor should take over. We conclude with identified directions for further work.

## I. INTRODUCTION

In automated driving, advanced sensors, functions and processing resources are required to deal with complex tasks and environments, with the context defined by so called Operational Design Domains (**ODD**'s). An "ODD" specifies the "conditions under which a given driving automation system or feature thereof is designed to function, including, but not limited to, driving modes", [1]; an ODD may thus e.g. specify environmental and speed restrictions, and a driving mode referring to e.g. high-speed cruising.

Driving automation requires to understand the location of the ego-vehicle and the operational situation (e.g. road ahead, and any obstacles such as people and other vehicles). It further requires to plan and take action for short and intermediate time horizons of the driving, referred to as operational and tactical functions part of the dynamic driving task (**DDT**), [1]. Accomplishing the DDT under uncertain and varying conditions, including mistakes/careless behavior of other road users and the many components in an automated vehicle that may fail, make higher *levels* [1] of automated driving challenging. It is intrinsically hard to conceive all situations that may arise a-priori, [2]. Since automated driving pushes the limits of existing technologies and methodologies, it is not surprising that there are not established best practices on building such high-performance systems that are sufficiently safe and reliable, while not prohibitively expensive.

Traditional engineering of safety critical systems mandates appropriate risk reduction through a number of process and product measures, see e.g. [3]. There is a strong tradition towards the use of simplicity for safety critical parts, and heavy redundancy when safety requires availability (such as in aircraft control systems), see e.g. [4]. We believe that the understanding of such traditional concepts in the context of high levels of driving automation is still insufficient.

The main functionality needed for automated driving, especially for the DDT, will be highly complex, in terms of algorithms, the use of machine learning systems with opaque behavior and difficulty in predicting failures [5], and in terms of its computing hardware and software. This main functionality will also clearly be safety critical since it controls the overall vehicle behavior. This means that the corresponding safety integrity level of ISO26262, [3], is likely to be the most severe one. With current safety engineering practices, it is hard to conceive how safety assurance could be provided. For example, while duplicating the main functionality (i.e. providing redundancy), may help to improve availability, it does not solve the issue of safety assurance. First, the safety assurance problem remains. Secondly, it is difficult to deal with common cause failures. Thirdly, if two diverse and actively redundant channels are provided, how to handle the situation when they do not agree?

The primary focus of this paper is therefore on the architecting of an Autonomous Driving Intelligence (**ADI**) [6], that attempts to addresses the limitations of the state of the art. With ADI we refer to the computational intelligence, sensors and communication needed for high levels of automated driving – thus on key functions for performing the DDT. We focus on level 3 of driving automation and above, [1], providing a problem formulation and analysis of the architecting problem in order to address concerns for costs, safety and availability. In particular, we present a functional architecture of a **fault-tolerant ADI that encompasses a nominal channel as well as a safety supervisor channel**, where the safety supervisor represents a redundant and simpler set of functionalities that should take over if the nominal channel fails. We also provide a formalization of the conditions of takeover by the supervisory channel in terms of the scenarios the ADI will encounter. Overall, our approach promises to facilitate safety assurance for the ADI as a whole and suggests ways of providing cost-efficient fault-tolerance.

We draw upon a multidisciplinary analysis of the problem and experiences from multiple case studies and industri-

This work is funded by ARCHER (2014-06260 Vinnova/FFI), AutoDrive (Grant Agreement No. 737469) and EUREKA TRACE (proj. No. CAT311).

al collaboration [6]–[9]. While aware of the limitations of functional safety standards for higher levels of automated driving, [10]–[12], we make use of established concepts and terminology as far as possible, including the ISO26262 [3].

Section II provides the overall problem formulation for a fault-tolerant ADI. In Section III we briefly review state of the art. In Section IV, we present our fault-tolerant ADI architecture, discuss the sources of hazardous events (what can go wrong), the division of responsibilities among the nominal and supervisor channels, and when the supervisor should take over. Finally, in section V, we provide a discussion and identify directions for further work.

## II. PROBLEM FORMULATION

We now present our problem formulation and the assumptions made. Our main focus is to investigate ADI behavior at the vehicle level.

According to the ISO 26262, [3], a *hazard* refers to a "*potential source of harm caused by malfunctioning behavior*" of a system, and/or a function, (ISO26262 here uses the term "item"). Further, a *hazardous event* refers to "*a combination of a hazard and an operational situation*".

Aligned with these definitions, we will use the term hazardous event to refer to a **system level state** – encompassing the state of the ego vehicle and its environment - that may lead to harm (in other words propagate to an accident).

In this paper we consider four sources of hazardous events: (i) **random hardware faults**, (ii) **systematic faults**, (iii) **performance limitations** of ADI functionalities, and (iv) **operational situations** (relating to the states of all entities outside the ADI including the vehicle environment).

Hazardous events caused by (i) and (ii) are the focus of ISO 26262 [3], referring to faults in the embedded system of the ego-vehicle. Performance limitations, (iii), such as the ability to detect surrounding objects in various conditions, may cause improper/erroneous internal representations such as false negatives, [10], [12], [13], and may thus lead to hazardous events. We note that such performance limitations only manifest in particular operational situations and that limitations will always be present to some degree. Operational situations, (iv), could for example be due to a careless or distracted driver of another car.

We thus conclude that hazards can be caused by vehicle internal (i, ii), external (iv), and combined, (iii) sources.

We assume that a vehicle is composed of an ADI and a platform that together perform the DDT. With platform we refer to the mechanical, electrical, electronic and software parts of a vehicle that provide basic functionalities such as steering, braking, propulsion, power, etc. We assume that the platform can largely be made fault-tolerant, i.e. able to compensate for errors in the steering system, possibly providing backup degraded performance. For some errors however, such as a flat tire, fault-tolerance cannot be assumed. We moreover assume that the platform is able to report its status to the ADI. The ADI comprises two channels. The nominal channel performs the DDT. The safety supervisor channel deals with situations when the nominal channel has *critical errors*, referring to sources (i) and (ii) of hazardous events, and potentially to (iii) (further elaborated in Section IV.B). The safety supervisor is designed to provide degraded ADI functionality, required to reach a minimal risk condition i.e. execute a safe maneuver within a limited time-frame. It also needs to be able to accomplish a variety of safety maneuvers (e.g. emergency stop, maneuver to the side of the road) in a sufficiently deterministic fashion to enable safety assurance.

Achieving safety corresponds to ensuring appropriate risk reduction, including the ability to achieve a *minimal risk condition* ([1]) after occurrence of a system failure, an ability required for levels 4 and 5 of automated driving. Achieving such a condition will require (at least) a degraded ADI capability, since simply stopping when a critical error is detected could be highly undesirable and risky in some situations (e.g. left lane on a high-way or in a tunnel).

Fig. 1 provides a simplified hierarchical state-machine model that depicts overall vehicle states, and how hazardous events may arise. The top-level super-state (with three sub-states) illustrates driving in operational situations without critical errors in the ADI of the ego-vehicle. The sources for hazardous events in this super-state rely on events at the traffic level (e.g. a dangerous maneuver by another vehicle), when the ODD is about to be exited, or by non-ADI errors such as a flat tire. Nominal operation of the ADI involves **M**onitoring the driving environment, situation **A**nalysis (e.g. classifying objects and assessing intentions), operational and tactical **P**lanning, and **E**xecution of these plans. This corresponds to elements of the classical **MAPE-K** control loop, where **K** corresponds to knowledge used in this process, such as models of vehicles and HD maps, [14]. The nominal operation also encompasses situations with intended (temporary) degraded performance, e.g. a case with direct sunlight on cameras – causing more conservative operation for a short period of time.

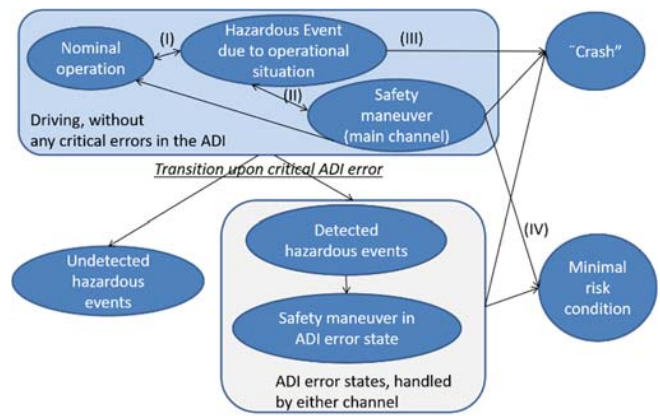

Fig. 1. State machine model illustrating sources of hazardous events.

The ego-vehicle thus constantly assesses risks and would typically aim to pro-actively minimize risks by proper planning at the tactical and operational levels, see e.g. [15], [16]. A sudden, unexpected maneuver by another vehicle may however cause a hazardous event (transition (I) in Fig. 1).

Unless it is detected in time, a crash may ensue (transition (III)). If the ADI on the other hand detects a hazardous event, it may decide to initiate a safety maneuver (transition (II)), with the purpose to transition to a minimal risk condition (transition (IV)).

The transition from the top-level super-state ("downwards" in Fig. 1) is taken when a critical error in the ADI occurs. It is clearly essential that such errors are detected with high probability to be able to keep risks at acceptable levels. If detection is not possible, however, the vehicle will enter undetected "hazardous events" with unknown consequences. This state could lead to a crash, later detection, or possibly to inherent recovery back to fault-free operation (for simplicity, these various transitions are not shown in Fig. 1).

If the nominal channel is able to detect critical ADI errors, it will perform a safety maneuver. Similarly, if the error is in a critical part of the nominal channel itself, such that it cannot identify or deal with the error, the safety supervisor channel should be able to take over control and perform a safety maneuver. It then also becomes essential that the interactions between the nominal and supervisor channels do not cause new types of failure modes, further discussed in Section V. Faults in the supervisor itself are discussed in Section IV.B.

### III. Related work

We here briefly review state of the art, positioning our work w.r.t. the fields of computer science, embedded systems, safety and fault-tolerant computer engineering, and perspectives from the state of practice in the industry.

*A. Safety monitors*

A safety monitor observes the behavior of the system of interest (SOI) and checks it against a set of predefined safety properties, which are derived from system requirement documents [17] or hazard analysis [18]. In the terminology used in this domain, monitoring may encompass a range of capabilities from corrective interventions to taking over full control. Safety properties are normally defined in natural language based on system states [17]–[19]. E.g. in [18], one safety property is defined as "the robot velocity should not exceed $v_0$ if its arm is un-folded". Once violations are detected, the safety monitor will initiate corrective [19]–[22] or preventive [18], [22] action during runtime. Safety properties are desirable to describe with formal languages such as first order logic [20] or temporal logic [17]–[19], [23]. Work on monitors have also considered various sources of uncertainties including, (1) observability restriction, and (2) violation of injective mapping from observation to the real states [17], [23]. Related to (1), limitations of observations due to discretization, quantization, and measurement errors are discussed in [17]. Commonly used techniques for dealing with uncertainty in the context of safety monitoring include approaches such as Bayesian method [24], and Dempster-Shafer Logic [2].

The concept of the safety monitor can be considered as a variant of the Simplex architecture [21], in which the behavior of a high-performance controller is monitored and where control is switched to a high assurance controller if necessary. It is assumed that a state space envelope can be determined at design time, and under which the high assurance controller will be able to take over control. This envelope can then be used to determine when the monitor should switch over to the high assurance controller. The application of Simplex architecture promotes the use of structured diversity in a high assurance control subsystem. Our work can be seen as an extension of the Simplex architecture, by employing structured diversity, but where we relax its "closed world assumption" and consider also alternative approaches to heavy redundancy for the high assurance channel.

*B. Safety and fault-tolerance engineering*

Safety engineering is concerned with methods to assess, eliminate and/or reduce risk to acceptable levels where risk is seen as the combination of probability of occurrence of harm and severity of that harm [3]. The core of most relevant safety standards (e.g. the ISO 26262 [3] and IEC 61508 [25]) lies in the identification of hazards and hazardous events that may occur during the operation of the system, identifying the risks they pose and the means to eliminate or mitigate them to acceptable levels.

Fault-tolerance has the purpose to use either application specific or systematic techniques to increase the reliability of a system, in particular such that it can tolerate a certain number and types of faults [4]. The dependability terminology by Avižienis et al. [26], uses the terms faults, errors and failures, where a fault is the cause of an error which represents the incorrectness of a state in the system. This may cause a failure in the system i.e. inability to deliver correct service – which in turn may result in a hazardous event.

Diversity such as the use of N-version programming, or diverse underlying hardware is common in fault-tolerant systems, requiring sufficient independency between the channels [3]. It is however also seen as controversial. For example, the FAA DO 178B discourages the use of N-version programming as a primary tool to achieve software reliability, since it is very hard to quantify its effects, [27]. Diversity per se is not guaranteed to enhance neither availability nor safety. However, a structured approach to diversity with anchoring in the architecture including recovery blocks and forward recovery techniques [26], [28], has been argued to be key for enhanced availability and cost-efficiency [21].

Automotive systems where a minimal risk condition is usually available in a time duration that is of much shorter duration than the mission time, can use a safety system to ensure the system is safe by compromising on the mission itself. Koopman et al. [5] suggest that this compromise of availability for the purposes of safety might be the first step in getting the vehicles into market.

While the risk of hazards caused due to the *performance limitations of intended functionality* can be limited to a certain extent by the limitation/modification of the ODD (e.g. a separate lane for automated vehicles, specific V2X infrastructure to provide additional information), there are practical limits to how much this can be achieved without compromising the utility of the function. A consequence of this, as discussed in [2], is that it may be necessary to adopt runtime risk assessment and run time verification techniques to achieve the reduction in risk. To this end, concepts such as monitors and constraints (section III.A) - (such as the definition of operational envelopes used in [29]) can be used to allow for the nominal channel to operate freely within a set of bounded conditions. The responsibility of detection of a potential violation of the bounds resides within the monitor leaving the nominal channel to be classified at lower ASIL levels. In this paper we further elaborate on the use of such monitors and the challenges in dealing with multiple sources of hazardous events (Section II).

*C. Industrial safety practices for automated vehicles*

In terms of the state of practice in the industry, the safety reports released by General Motors [30] and Waymo [31], show the incorporation of several of the techniques discussed in this section. Common themes of redundancies include duplication of all critical subsystems and the use of diversity. Diversity is for example used for the independent collision avoidance systems and for perception in conjunction with the overlap of the sensory fields of view. Little detail is however given for how various types of faults (such as those we described in Section 2) are dealt with and how inconsistencies between the duplicated channels are resolved.

IV. A FAULT-TOLERANT FUNCTIONAL ARCHITECTURE FOR AUTOMATED DRIVING

We here address topics related to the architectural design of a fault-tolerant ADI, including, (i) high-level design principles (this Section), (ii) structural design of the fault-tolerant ADI (Section IV.A), (iii) division of responsibilities among the channels of the fault-tolerant ADI (Section IV.B), and (iv) safety constraints which describe the conditions when the safety supervisor should take over control from the nominal channel, (Section IV.C).

In developing the fault-tolerant ADI, we have adopted a number of principles that have guided our design choices:

- P1: No single point failures in the ADI.
- P2: Failures of the supervisor must be handled by transitioning to a minimal risk condition.
- P3: The design should support instantiations to technical architectures that in various ways trade-off cost, safety and availability.
- P4: Determinism and transparency support safety assurance in the supervisory channel. Achieving transparence favors the use of first principles (model-) based algorithms. This also implies that we aim for diverse solutions (recall end of section III.A).
- P5: Common cause failures between the nominal and supervisor channels should be avoided. A common mode failure could otherwise cause both channels to fail at once.
- P6: Coherence in design among the two channels should be emphasized so that the safety supervisor only takes over control when necessary, and that it does not miss taking over control when needed.

Actual availability and reliability requirements for a fault-tolerant ADI will depend on the intended vehicle usage, and the desired level of vehicle availability.

*A. A functional architecture for automated driving*

As depicted in Fig. 2., the ADI is composed of two channels, the nominal and the safety supervisor (for simplicity the latter is in the following referred to as supervisor for short). Both channels contribute to automated driving safety.

Each channel incorporates MAPE-K loops with corresponding 5 components, **M**, **A**, **P**, **E** and **K**, where a superscript refers to the channel number, with "1" referring to the nominal channel, and "2" and "3" referring to the supervisor ("2" vs. "3" explained in the following). These loops are briefly elaborated in the following:

**Nominal channel (*Nc*)**: The MAPE-K loop of the *Nc* is depicted in Fig. 2. We note that the *Nc* includes a world model – corresponding to the $K^1$-component, incorporating models of for example object's behaviors and maps, and a metric or function for assessing risk dynamically. Using these models and sensory observations from $M^1$, the $A^1$ will detect and localize objects, predict their near-term behavior, and based on this assess the current risk. This enables to determine dynamic driving envelopes which are considered to have acceptable risk. The envelopes will evolve dynamically as the ego-vehicle and other object moves, and as new static and moving objects appear, [15], [16]. The $P^1$ component is usually comprised of three functions. A behavior decision function selects an appropriate driving behavior for the current traffic scene, e.g., "continue in current lane" or "change lane left". Accordingly, the trajectory planning function generates a reference trajectory avoiding static and dynamic obstacles and subject to vehicle dynamic constraints. Motion control functionality computes appropriate steering, throttle and brake input such that the reference trajectory is precisely followed. The platform is considered as a shared resource that provides the **E** (execution) component for both (nominal and supervisor) channels.

**Supervisor channel (*Sc*)**: The *Sc* incorporates two MAPE-K loops, where one (labelled with superscript notation $^3$) has the special role to detect critical errors in *Nc*, and switch the control from the *Nc* to the *Sc* when such errors are detected. Control is then handed over to the *Sc* MAPE-K loop with superscript notation $^2$, referring to functionality that corresponds to a degraded version of the *Nc*.

$M^2$ is shared among the two MAPE-K loops of *Sc*. $A^3$ assesses internal and external monitored states, and $P^3$ is responsible for the critical decision for taking over control – triggering the switch to remove the nominal channel from control (see control flow IX in Fig. 2.). Once the control has been switched, the $M^2A^2P^2$ components will control the vehicle (i.e. monitor, analyze and plan for the execution). We remark that $A^2P^2$ – as responsible for continued situation analysis, safe maneuver planning and choice of maneuver – are always active *during runtime*, such that the supervisor is readily available to take over once the switch occurs. $K^{23}$ for the *Sc* includes a simplified world model as well as "safety constraints" which for the *Sc* define the conditions when it should take over control from the *Nc* (relating to principle P4). For simplicity the connection of **K** to other components within the corresponding channel are not shown explicitly.

The interactions between the platform, the *Sc* and the *Nc* are illustrated in Fig. 2. The platform provides a set of states to both the *Nc* and *Sc* (flow IV). These states describe the actual state and capabilities of the platform in terms of nominal or degraded modes. The *Sc* optionally (depending on the configuration) – provides a "live signal" to the *Nc* (flow VIII). This allows a potentially cost-efficient way of dealing with hardware failures of the supervisor channel (referring to principles P2 and P3). As an option, the supervisor channel could be configured to be redundant (P3). The *Nc* communicates 3 sets of states to the *Sc* (flows I, II and III):

- Flow I: Status of sensory systems and, depending on the perception system configuration, also access to some sensory data (e.g. objects from smart sensors) as well as the self-diagnostic information of the *Nc*.
- Flow II: Nearby objects and their expected intents.
- Flow III: Near term execution plan including intent of maneuver (e.g. a safety maneuver)

We consider that the *Sc* should have an additional set of sensors that are independent to the *Nc*. These sensors need to be highly reliable for use in the short-term safety maneuvers of the *Sc*. They could be seen as an evolution of today's sensors used in active safety systems.

### B. Division of responsibilities – dealing with failures

Since both channels contribute to automated driving safety, the question arises about division of responsibilities. The main responsibility of the *Nc* is to accomplish the DDT and deal with hazardous events due to the operational situations within the given ODD, originating from other objects or from the platform of the ego-vehicle. The main responsibility of the *Sc*, on the other hand, is to monitor and analyze whether the *Nc* is healthy, and to provide back-up functionality and risk mitigation when needed. Translating this to the sources of hazardous events discussed, yields the following division of responsibilities, as illustrated in TABLE I. We use **X** to denote responsibility for run-time detection and handling, and the right-most column identifies needed design-time measures. **Y** is used to denote a responsibility which requires further investigation.

Referring to the four sources of hazardous events described in section II, the *Sc* is designed to handle systematic and random hardware faults of *Nc*. Questions remain whether *Sc* can also mitigate hazardous events caused by external events and performance limitations of *Nc*. Using simpler and more defensive perception for *Sc* might help to address certain performance limitations of *Nc* but may also cause new types hazardous events, e.g. caused by false positives. Therefore, it is reasonable to assume that the *Sc* cannot overcome all the performance limitations of *Nc*, or at least that this is non-trivial and requires further investigation (see Section V). As mentioned previously, the *Nc* will monitor the *Sc*, requiring that the *Sc* has its own error detection mechanisms.

### C. Safety constraints and interactions between the channels

Safety constraints determine when the *Sc* should take over from the *Nc*. Safety constraints can be divided into two sets: *internal* safety constraints (ISCs) that are defined using the status of the *Nc* and the vehicle platform, and *external* safety constraints (ESCs) that are defined with respect to driving scenarios. According to TABLE I., the *Sc* can only take over when critical errors of the *Nc* are detected.

ISCs describe for which critical errors of the *Nc* and the vehicle platform, the *Sc* shall take over. The detection of *Nc* errors is facilitated if the *Sc* has access to internal states of the *Nc*, which include the communication ports of the software components (referring to the flows I, II and III in Fig. 2.) as well as status and diagnostic information (e.g., error flags/reporting from online diagnosis or watchdogs, referring to flows I and IV). If self-diagnosed errors are reported from the *Nc*, the *Sc* will directly take over. Besides listening to the error reporting, the *Sc* will also check the behavior of the safety critical functions ($M^1$-$A^1$-$P^1$) in the *Nc* against their expected behavior through techniques like runtime verification and plausibility checks. *Sc* may for example check the deviation between the intended trajectory of the ego-vehicle, generated by the trajectory planner of the *Nc*, with the actual trajectory observed by the *Sc*.

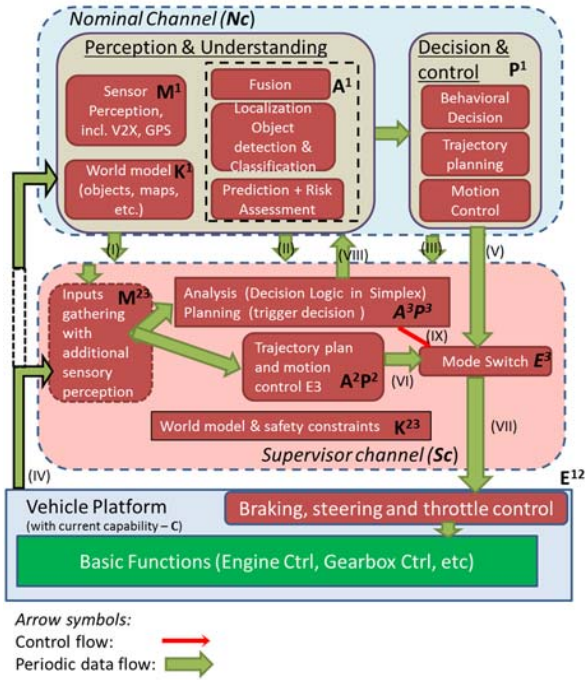

Fig. 2. A functional architecture for an ADI for high levels of driving automation. (Meanings of the flows: (I): Sensor status, *Nc* diagnostic states; (II): Objects, pos/vel etc. (III): Near term trajectory; (IV) Platform status and diagnostic states; (V): *Nc* set-points for platform; (VI): *Sc* set-points for platform; (VII) Set-points for platform; (VIII): *Sc* live signal; (IX): Control flow to mode switch)

TABLE I.   DIVISION OF RESPONSIBILITIES IN DEALING WITH FAULTS/FAILURES (DESIGN TIME AND OPERATIONAL CHANNELS)

| Sources for Hazardous Events | Run time: Detection responsibility | | Design time measures |
|---|---|---|---|
| | Nc | Sc | |
| External Sources | X (detected by A[1]) | - | Determining risk estimation functions |
| Faults in the platform | X (detected by A[1]) | - | Design to detect and handle |
| Random HW faults in *Nc* | X (internal error detection) | X (through safety constraints) | Ensure independence. |
| Random HW faults in *Sc* | X (detection by absent *Sc* live signal) | X (internal error detection) | Configuration according to actual requirements |
| Systematic faults in *Nc* | X (internal error detection) | X (through safety constraints) | Eliminated as far as practically possible |
| Systematic faults in *Sc* | X (detection by absent *Sc* live signal) | X (internal error detection) | Risk reduction through best safety practices |
| Common Cause faults | - | - | Best safety practices; efforts to avoid negative interactions among *Nc* and *Sc* |
| Performance limitations in *Nc* | Y | Y (through safety constraints) | Potential design of active safety like functionality as part of *Sc*, e.g. to deal with a false negative in perception by the *Nc* |
| Performance limitations in *Sc* | Y | Y | Design to minimize risk of false positives and negatives. |

Due to the complexity of the *Nc*, complete error detection can hardly be guaranteed. Therefore, ESCs also need to be monitored to guarantee safety on the vehicle level.

ESCs describe the boundary of acceptable driving scenarios. They are normally derived from system requirements [17] or hazard analysis [18], and will be used for the design of both the *Nc* and the *Sc*. Each driving scenario consists of the behavior of the ego-vehicle and also the environment. According to the theory in [17], each driving scenario can be defined using relevant system level states $q_1, q_2, ..., q_n$ of types $Q_1, Q_2, ..., Q_n$, with $St = Q_1 \times Q_2 \times ... \times Q_n$ denoting the physical states of the ego vehicle and its environment. These states can be further classified into two sets:

- The controlled states, $q_1, q_2, ..., q_i$, i.e., the behavior of the ego-vehicle (e.g., position, speed and acceleration. Uncontrolled states include the behaviors of all the traffic users, the road condition and the traffic information).

- The monitored states, $q_j, q_{j+1}, ..., q_n$, i.e., the ones that are monitored by the perception and localization systems (i.e. components M1 and M2 in Fig. 2.). They reflect the observed driving scenario. Most of the controlled states are also monitored. One example of controlled but not monitored states can be the lighting of the vehicle.

The dynamics of these states are modelled as functions of time. A particular driving scenario can be represented by a function pair as $d^t = (m^t, c^t)$, where $m^t$ and $c^t$ are state functions of time $t$ with $m^t: \mathbb{R} \to Q_j \times Q_{j+1} \times ... \times Q_n$ defining the behavior of the monitored environment, and $c^t: \mathbb{R} \to Q_1 \times Q_2 \times ... \times Q_i$ defining the behavior of the ego-vehicle. $d^t(t)$ denotes the particular values of the states at time $t$, while $d^t$ refers to the corresponding state function. If $t$ is given as the ending time of the driving scenario, $d^t$ denotes all the previous states (i.e., a trajectory on $St$) before $t$. All the possible driving scenarios are represented by the set $D = \{d_1^t, d_2^t, ...\}$. Therefore, the ESCs determine the boundary of the set $ESC \subseteq D$ of all the safe driving scenarios, while unsafe driving scenarios (i.e., the hazardous events) are the ones outside the safety constraints, denoted as $\overline{ESC} = D \setminus ESC$.

Due to the inertia of the vehicle dynamic, and the stochastic nature of an uncontrolled environment, a hazardous event ($hd^t \in \overline{ESC}$) should be defined based on risk (referring to severity and probability). Examples of $hd^t$ include driving too close to the vehicle in front or changing to a lane with an approaching vehicle. Therefore, ESCs can be defined as $ESC = \{d^t | \hat{R}(m^t, \Delta t) < R_{max}\}$, where $\hat{R}(m^t, \Delta t)$ denotes the estimated safety risk within the coming $\Delta t$ seconds based on $m^t$, $R_{max}$ is the risk threshold and $\Delta t$ is considered as a reasonable time to react. As mentioned before, most of the $hd^t \in \overline{ESC}$ are derived during design-time. However, it is impossible to predefine all the hazardous events, leading to the need to perform $\hat{R}(m^t, \Delta t)$ at run-time (in both *Nc* and *Sc*). If the estimated risk exceeds $R_{max}$ (referring to the state "hazardous event due to operational situation" in Fig. 1), a safety maneuver (in *Nc*) should be triggered.

There are two causes of the violations of ESCs: hazardous events due to an operational situation (e.g., unsafe behavior of other vehicles) and hazardous events due to the (propagation of) critical *Nc* errors. Ideally, the first cause should be handled by triggering the safety maneuver of the *Nc* and the second cause should be detected earlier as a violation of ISCs.

Therefore, when driving in a known operational situation (referring to the nominal operation state in Fig. 1), control decisions made by the *Nc* should always guarantee that $d^t \in ESC$, while the *Sc* checks $m^t$ against $\overline{ESC}$. If $m^t \in \overline{ESC}$ is detected, it means an undetected error occurred in the *Nc* and the *Sc* shall take over. When driving in a hazardous operational situation with $\hat{R}(m^t, \Delta t) \geq R_{max}$, the *Nc* should perform its safety maneuver, If the safety maneuver (*Nc*) is not performed or ISCs are violated during the safety maneuver (*Nc*), the *Sc* shall take over.

To this end, the *Sc* needs to access both $St$ (the environment/"physical" states) to monitor ESCs and the internal states of the nominal channel (the "cyber" states) to monitor ISCs. It is clearly a challenge to ensure full coverage (completeness) of ISCs and ESCs. The incompleteness of ISCs can be at least partly compensated by the monitoring of ESCs. The incompleteness of ESCs can be compensated by a generic $\hat{R}(m^t, \Delta t)$ calculated at runtime. The ISCs and ESCs thus complement each-other and combining them should improve the error detection coverage.

## V. Discussion and directions for future work

We have presented a proposal for a fault-tolerant ADI that relies on structured diversity. There are several possible variations of the proposed architecture. For example, the configuration of the *Sc* may include various levels of redundancy depending on the achievable error detection coverage, cost constraints and availability requirements. The safety and availability requirements will in turn depend on the ODD, determining the feasibility of these options (for example the light weight redundancy option where the *Sc* provides a "live signal" to the *Nc*).

Even though both faults and performance limitations of the ADI may propagate to hazardous events, they are intrinsically different and should be treated in different ways during safety analysis. ISO 26262 [3] assumes that faults can be identified, fixed or tolerated (via e.g., independent redundancy). Performance limitations, on the other hand, will always remain to some extent, but can be mitigated by sensor redundancy/diversity, fusion, tracking, etc. (see e.g. [7]). Faults and performance limitations can both propagate and cause hazardous events.

During hazard analysis and risk assessment (HARA), ISO 26262 [3] assumes that faults are independent of operational situations. However, manifestations of performance limitations are commonly dependent on the operational situations. In other words, the exposure of the performance limitation likely implies the exposure of the corresponding hazardous event. In addition, machine learning algorithms are generally inscrutable (hence difficult for code review and white-box testing) and inductive (hence no formal requirements for formal verification and requirement-based testing) [5]. Therefore, the safety engineering methodology proposed by ISO 26262 [3] is not sufficient for dealing with hazardous events due to performance limitations of intended functions.

A key takeaway from our work is the difficulty in formulating and deriving safety constraints, as conditions or risk measures that cover all relevant hazardous events. Dealing with hazardous events stemming from both the operational situation and critical ADI errors makes design difficult. With a well-controlled ODD, many traffic related hazardous events will be unlikely or disappear, implying that it may be possible to simplify the design of the fault-tolerant ADI and that it will be easier to define safety constraints.

The design space and "fault" space (sources of hazardous events) to be considered in designing the fault-tolerant ADI is large. We have had many lengthy discussions on which functionality should be assigned where and to find a coherent division of responsibilities. Establishing and reasoning about confidence in perception is a challenging topic on its own.

We see this work as representing both a position paper and problem formulation. There are consequently many avenues for further work.

Design, realization and evaluation of the fault-tolerant ADI is a natural continuation that we plan to pursue, both through modelling/simulation and real implementation with experiments. Supervisor development requires the design of its core functionalities, i.e. realizing the components of a minimalistic and verifiable, yet sufficiently capable (degraded performance) MAPE-K control loop, inevitably involving trade-offs. Further work is needed to address the error coverage of the combined ISC and ESC mechanisms, suitable risk estimation functions, as well as potential causes for common mode failures between the *Nc* and the *Sc*. This also requires consideration of various uncertainties that the *Sc* has to deal with. Realization of the technical architecture of the ADI requires considering appropriate levels of redundancy and centralized vs. decentralized realizations of the supervisor – and corresponding analysis of availability and reliability. As one extension, one could envision that the supervisor could leave control back to the *Nc*. It is also of interest to investigate if the architecture can be extended to further exploit the redundancy among the channels – for example to enhance availability. The **Nc** will likely be equipped with a variety of diverse and redundant sensors. It would be relevant to investigate if and how the fault-tolerant ADI may have the *Sc* to make use of a subset of these for safety maneuvers. This relates to the topic of whether/how the *Sc* can mitigate the hazardous events caused by the performance limitations.

Another interesting direction is to evaluate and elaborate the proposed functional architecture towards a functional safety concept, [3]. The architecture that we have introduced is a first step towards a functional safety concept by describing safety measures and mechanisms as part of the architecture. To conform to a functional safety concept, safety goals and safety requirements need to be elaborated, and the feasibility of the approach proposed by ISO26262 has to be investigated (given that ISO26262 was not developed with automated driving in mind). It is clear that the supervisor as a whole, as well as its components/functionalities, will be highly critical. Stringent risk reduction measures need to be applied.

In contrast to a functional safety concept, however, we have considered system safety – in terms of a broader set of sources of hazardous events. A further step is to extend this fault model to incorporate cybersecurity considerations into the architecture design.

Finally, while we have presented work towards architectural solutions, this paper also highlights the complexity inherent in automated driving and the fact that the architecture alone will not suffice, (refer to TABLE I). It will be imperative for automotive organizations to emphasize safety engineering and complexity management, in terms of architectures, processes and organizations [32].